\documentclass[11pt,a4paper,fleqn]{article}
\catcode`\@=11
\@addtoreset{equation}{section}
\def\theequation{\arabic{section}.\arabic{equation}}
\def\appendix{\renewcommand{\thesection}{\Alph{section}}\setcounter{section}{0}
             \renewcommand{\theequation}
           {\mbox{\Alph{section}.\arabic{equation}}}\setcounter{equation}{0}}

\def\maketitle{\thispagestyle{empty}\setcounter{page}0\newpage
               \renewcommand{\thefootnote}{\arabic{footnote}}
                 \setcounter{footnote}0}
\renewcommand{\thanks}[1]{\renewcommand{\thefootnote}{\fnsymbol{footnote}}
              \footnote{#1}\renewcommand{\thefootnote}{\arabic{footnote}}}
\newcommand{\preprint}[1]{\hfill{\sl preprint - #1}\par\bigskip\par\rm}
\renewcommand{\title}[1]{\begin{center}\Large\bf #1\end{center}\rm\par\bigskip}

\renewcommand{\author}[1]{\begin{center}\Large #1\end{center}}
\newcommand{\address}[1]{\begin{center}\large #1\end{center}}
\newcommand{\pacs}[1]{\smallskip\noindent{\sl PACS numbers:
                      \hspace{0.3cm}#1}\par\bigskip\rm}
\def\babs{\hrule\par\begin{description}\item{Abstract: }\it}
\def\eabs{\par\end{description}\hrule\par\medskip\rm}
\renewcommand{\date}[1]{\par\bigskip\par\sl\hfill #1\par\medskip\par\rm}

\newcommand{\ocha}[1]{${}^{(#1)}$ Leading Graduate School Promotion Center,
Ochanomizu University, 2-1-1 Ohtsuka, Bunkyo-ku, Tokyo 112-8610, Japan\\}
\newcommand{\ochasci}[1]{${}^{(#1)}$ Faculty of Science, 
Ochanomizu University, 2-1-1 Ohtsuka, Bunkyo-ku, Tokyo 112-8610, Japan\\}
\newcommand{\dip}[1]{${}^{(#1)}$ Dipartimento di Fisica, Universit\`a di Trento\\
                                     via Sommarive 14, 38123 Trento, Italia\\}
\newcommand{\infn}[1]{${}^{(#1)}\,$TIFPA (INFN), Trento, Italia\\ \medskip}
\newcommand{\csic}[1]{${}^{(#1)}\,$ICE (CSIC-IEEC), UAB Campus, E-08193 Bellaterra, Barcelona \\}
\newcommand{\icrea}[1]{${}^{(#1)}\,$ICREA, Barcelona, Spain \\ \medskip}
\newcommand{\tomsk}[1]{${}^{(#1)}\,$Tomsk State Pedagogical University, 634061 Tomsk and \\
National Research Tomsk State University, 634050 Tomsk, 
Russia \\ \medskip}

\newcommand{\bamba}[2]{Kazuharu Bamba${}^{#1,#2}$\thanks{e-mail:\sl bamba.kazuharu@ocha.ac.jp\rm}}
\newcommand{\guido}[2]{Guido Cognola${}^{#1,#2}$\thanks{e-mail:\sl cognola@science.unitn.it\rm}}
\newcommand{\sergio}[2]{Sergio Zerbini${}^{#1,#2}$\thanks{e-mail:\sl zerbini@science.unitn.it\rm}}
\newcommand{\sergei}[2]{Sergei D. Odintsov${}^{#1,#2,g}$\thanks{e-mail:\sl odintsov@ieec.uab.es\rm\rm}}

\newcommand{\Eqn}[1]{&\hspace{-0.2em}#1\hspace{-0.2em}&}

\renewcommand{\ss}[1]{\subsection{#1}}

\def\segue{\qquad\Longrightarrow\qquad} 
\def\hs{\qquad}               
\def\nn{\nonumber}            
\def\beq{\begin{eqnarray}}    
\def\be{\begin{eqnarray}}
\def\eeq{\end{eqnarray}}      
\def\ee{\end{eqnarrayn}}
\def\ap{\left.}               
\def\at{\left(}               
\def\aq{\left[}               
\def\ag{\left\{}              
\def\cp{\right.}              
\def\ct{\right)}              
\def\cq{\right]}              
\def\cg{\right\}}             

\def\R{{\hbox{{\rm I}\kern-.2em\hbox{\rm R}}}}   
\def\H{{\hbox{{\rm I}\kern-.2em\hbox{\rm H}}}}   
\def\N{{\hbox{{\rm I}\kern-.2em\hbox{\rm N}}}}   
\def\C{{\ \hbox{{\rm I}\kern-.6em\hbox{\bf C}}}} 
\def\Z{{\hbox{{\rm Z}\kern-.4em\hbox{\rm Z}}}}   
\def\ii{\infty}                                  
\def\Tr{\mathop{\rm Tr}\nolimits}                  
\def\Res{\mathop{\rm Res}\nolimits}                
\renewcommand{\Re}{\mathop{\rm Re}\nolimits}       
\def\dir{/\kern-.7em D\,}                          
\def\lap{\Delta\,}                                 
\def\al{\alpha}
\def\ga{\gamma}
\def\de{\delta}
\def\ep{\varepsilon}
\def\ze{\zeta}

\def\ka{\kappa}
\def\la{\lambda}
\def\ro{\varrho}
\def\si{\sigma}

\def\ph{\varphi}

\def\Ga{\Gamma}

\def\La{\Lambda}

\def\be{\begin{equation}}
\def\ee{\end{equation}}
\def\bea{\begin{eqnarray}}
\def\eea{\end{eqnarray}}

\def\nn{\nonumber}

\renewcommand{\title}[1]{\begin{center}\Large\bf #1\end{center}\rm\par\bigskip}
\renewcommand{\author}[1]{\begin{center}\Large #1\end{center}}

\begin{document}

\preprint{OCHA-PP-322}

\title{One-loop Modified Gravity in de Sitter Universe, Quantum 
Corrected Inflation, and its Confrontation with the Planck Result}
\author{\bamba{a}{b}, \guido{c}{d}, \sergei{e}{f} and \sergio{c}{d}}
\address{\ocha{a}\ochasci{b}\dip{c}\infn{d}\csic{e}\icrea{f}\tomsk{g}}

\begin{abstract}
Motivated by issues on inflation, a generalized modified gravity model is 
investigated, where the model Lagrangian is described by a smooth function 
$f(R, K, \phi)$ of the Ricci scalar $R$, the kinetic term $K$ of a scalar field $\phi$. In particular, the one-loop effective action in the de Sitter background is examined on-shell as well as off-shell in the Landau gauge. 
In addition, the on-shell quantum equivalence of $f(R)$ gravity in the Jordan and Einstein frames is explicitly demonstrated. 
Furthermore, we present applications related to the stability of the de Sitter solutions and the one-loop quantum correction to inflation in quantum-corrected $R^2$ gravity. 
It is shown that for a certain range of parameters, the spectral index of the curvature perturbations can be consistent with the Planck analysis, 
but the tensor-to-scalar ratio is smaller than the minimum value within the 
1 $\sigma$ error range of the BICEP2 result. 
\end{abstract}


\section{Introduction}\label{intro}

Relativistic theories of gravity have attracted a lot of interests in modern cosmology after the discovery of the current cosmic acceleration, namely, the dark energy problem, as well as after the confirmation of the early-time inflationary era. There exist several possible descriptions of the current accelerated expansion of the universe. The simplest one is the introduction of the small positive cosmological constant in the framework of General Relativity, so that we can deal with a perfect fluid whose equation of state (EoS) parameter $w=-1$. This fluid model is able to realize the current cosmic acceleration, but also other kinds of fluid (e.g., phantom, quintessence, inhomogeneous fluids...) with their suitable EoS have not been excluded yet (for a recent review, see~\cite{bamba}), because the observed small value of the cosmological constant leads to several conceptual problems, such as the vacuum energy and the coincidence problem.
For this reason, several different approaches to the dark energy issue have been proposed. 
Among them, the modified theories of gravity represent an interesting extension of General Relativity (for example, see~\cite{review} and references therein).

On the other hand, very recently, after the release of the Planck mission results~\cite{Ade:2013lta, Ade:2013uln}, a lot of papers concerning the inflationary era have appeared. Among many models of inflation proposed in the past years, only a restricted class of models based on a single scalar field theory seems to be in agreement with the Planck data. In particular, the so-called Starobinsky inflation~\cite{staro}, where the action consists of 
the Einstein-Hilbert term plus a quadratic term in the Ricci curvature (i.e., $R^2$ gravity), seems to be quite successful. Note that the gravitational sector 
of the Starobinsky inflation model is equivalent to that of the scalar inflationary model proposed in Ref.~\cite{sha}. Recently, a flow of works have been executed~\cite{recent}, in which such inflationary models related with $R^2$ gravity and its generalizations have been investigated. 

$R^2$ gravity is one of the simplest modified gravity models. 
For this reason, it may be of interest to explore generalized modified gravitational models represented by the Lagrangian involving an arbitrary dependence on the Ricci scalar $R$, a scalar field $\phi$ and its kinetic energy $K$.

In this paper, 
we clearly show the on-shell quantum equivalence of $f(R)$ gravity in the Jordan and Einstein frames. 
It may also be interesting to compute the one-loop effective action of such a generalized model in a maximally symmetric space, namely, the de Sitter space, which is one of the most relevant ingredients due to its applications 
to inflation in quantum-corrected $R^2$ gravity. 
This kind of computation has been performed for $f(R)$ gravity in Ref.~\cite{cogno05}. Here, our aim is to extend it in these generalized modified models. The results of such a study may directly be applied to the realization of the inflationary epoch by taking account of quantum gravity corrections. The one-loop effective action is calculated in the Euclidean sector, that is, the de Sitter space becomes the four dimensional compact sphere $S_4$ and the evaluation of the several functional determinants is made by making use of the zeta-function regularization (see, e.g.,~\cite{report}), and making use of the quantum field theory (QFT) in curved space-time~\cite{buch, BD-MW}. 
Typically, the so-called on-shell one-loop effective action is relatively simple to compute, and hence it may be used in order to study the stability of the generalized background and the on-shell quantum equivalence between the Jordan and Einstein frames. 
Instead, the off-shell one-loop effective action suffers from gauge ambiguities, and therefore in order to avoid the problem, one has to evaluate it in the so-called Landau gauge~\cite{frad}, which is a somehow selected gauge owing to the relation with the gauge-fixing independent description. The off-shell one-loop effective action is useful for examining the relevance of the quantum corrections to inflation observables. 
We note that in Refs.~\cite{K-T-V, B-E}, 
quantum corrections with completely different approaches from our method 
have been discussed. 

In addition, we derive the spectral index $n_\mathrm{s}$ 
and the tensor-to-scalar ratio $r$ so that we would examine 
whether those values can be consistent with not only 
the Planck data~\cite{Ade:2013lta, Ade:2013uln} 
but also the BICEP2 result~\cite{Ade:2014xna} on 
the $B$-mode polarization of the cosmic microwave background (CMB) radiation. 
One-scalar and quantum-corrected inflationary models consistent with 
the BICEP2 data have recently been discussed in Ref.~\cite{BICEP2-SI-QCI}. 
As a proposal for inflation in $R^2$ gravity, the trace-anomaly driven 
inflation in $f(R)$ gravity~\cite{Bamba:2014jia} has been studied, and 
the $n_\mathrm{s}$ and $r$ in it have been compared with 
the Planck and BICEP2 observations. 
We adopt the units $k_{\mathrm{B}} = c = \hbar = 1$ and express the 
Newton's constant as  
$G=1/M_\mathrm{P}^2$ with $M_\mathrm{P} =2.43 \times 10^{18}$ GeV the reduced Planck mass. 

The organization of the paper is the following. 
In Sec.~II, the generalized modified gravity model is introduced and the classical equation of motion is derived.
In Sec.~III, the one-loop quantization of the generalized gravity model is presented in detail. 
In Sec.~IV, the two one-loop effective actions, the on-shell and off-shell ones are written down in terms of functional determinants. 
In Sec.~V, cosmological applications are presented. Particularly, the quantum corrections to the Starobinsky inflation are evaluated in the limit of large scalar curvature. Moreover, we examine the spectral index of scalar modes of the density perturbations and those tensor-to-scalar ratio, and investigate whether our model can explain the observational consequences found by the Planck satellite 
and the BICEP2 experiment. 
The paper ends with the Conclusion and two Appendices, where technical details are given. 

\section{Generalized Modified Gravity Models}

The modified gravity model which we are interested in 
is described by the action 
\beq
I=\frac1{2\ka^2}\,\int d^4x \sqrt{-\tilde g} f(\tilde R,\tilde K,\tilde \phi) \,,\hs\hs
\ka^2=8\pi G\,.
\label{action0}
\eeq
The Lagrangian density $f$ is a smooth function depending on the Ricci scalar 
$\tilde R$, 
the scalar field $\tilde \phi$ and the kinetic energy 
$\tilde K=(1/2)\,
\tilde g^{ij}\tilde\nabla_i\tilde\phi\tilde\nabla_j\tilde\phi$, 
$\tilde g_{ij}$ is the metric tensor, $\tilde g$ its determinant, and 
$\tilde\nabla_i$ the covariant derivative, where $i,j,k,...$ are tensorial indices that run over the range $0, \dots,3$. 
Here and in the following, we use the tilde for arbitrary quantities. 

The field equations may be obtained by making the functional variation of 
the action in Eq.~(\ref{action0}) with respect to the metric
$\tilde g_{ij}$ and the scalar field $\tilde \phi$, given by 
\beq
\ag\begin{array}{l}
f_{\tilde R}\,\tilde R_{ij}-\frac{1}{2} f\,\tilde g_{ij}
       +\at\tilde g_{ij}\tilde\lap-\tilde\nabla_i\tilde\nabla_j\ct f_{\tilde R}
          +\frac12\,f_{\tilde K}\,\tilde\nabla_i\tilde\phi\tilde\nabla_j\tilde\phi=0\,,\\
\tilde g^{ij}\tilde\nabla_i\at f_{\tilde K}\,\tilde\phi\,\tilde\nabla_j\phi\ct=f_{\tilde \phi}\,,
\end{array}\cp
\label{Eeq1}
\eeq
where $\tilde\lap$ is the covariant Laplace operator in the metric $\tilde g_{ij}$, while 
$f_{\tilde R}$ represents the derivative of $f(\tilde R,\tilde K,\tilde\phi)$
with respect to the variable $\tilde R$ 
($f_{\tilde K}$ and $f_{\tilde \phi}$ have the similar meanings). 

{}From equations (\ref{Eeq1}), we see that there exists a constant curvature
solution $\tilde R=R$. With a constant scalar field $\tilde\phi=\phi$, we 
find that the Lagrangian density satisfies the following conditions:
\beq
f_R\,R_{ij}-\frac12\,f_0\,g_{ij}=0\,,\hs\hs  f_\phi=0\,,
\label{onShell}
\eeq
where all the quantities in the latter equation are evaluated on the solution $\{R,\phi\}$. In particular, $f_0=f(R,K,\phi)$.
In such a case, the background fields (constant scalar field and curvature) 
are solutions of the equations
\beq
R\,f_R-2f_0=0\,,\hs\hs f_\phi=0\,.
\label{BGF}
\eeq
Using these equations, 
a number of classical constant curvature solutions may be constructed.

\section{One-loop quantization around a maximally symmetric solution}

In this section, we discuss the one-loop quantization of the 
model on a maximally symmetric space. 
In these investigations, as quite usual, 
this should be regarded as only an effective approach 
(see, a book~\cite{buch}). 
We may start from the Euclidean action
\beq
I_\mathrm{E} [\tilde g,\tilde\phi]=-\frac{1}{16\pi G}\int\:d^4x\,\sqrt{-\tilde g}\,f(\tilde R,\tilde K,\tilde\phi)\,,
\eeq
where the generic function $f$ satisfies --on shell-- 
the conditions in Eq.~(\ref{onShell}), that 
ensure the existence of the solutions of a 
constant curvature $R$ and a constant scalar field $\phi$. 
Here, we are interested in, particularly, $S^4$ (de Sitter) case, 
but also $H^4$ (anti de Sitter) or $\R^4$ (euclidean) cases 
are included in the general discussion. 
In all such cases, we have 
\beq
R_{ijrs}=\frac{R}{12}\at g_{ir}g_{js}-g_{is}g_{jr}\ct \:, \hs
R_{ij}=\frac{R}{4}\,g_{ij}\,, 
\hs 
R=constant\,, 
\label{AAA2} 
\eeq
where $g_{ij}$ is the metric of the maximally symmetric space. 
In the next step, 
we examine the small fluctuations around the constant curvature solution, 
that is, 
\beq
\ag\begin{array}{ll}
\tilde g_{ij}=g_{ij}+h_{ij}\:, & |h_{ij}|\ll1\,,\\
\tilde g^{ij}=g^{ij}-h^{ij}+h^{ik}h^j_k+{\cal O}(h^3)\:, & h=g^{ij}h_{ij}\:,\\
\tilde\phi=\phi+\ph\,, & |\ph|\ll1\,, 
\end{array}\cp
\eeq
where indices are made lowered and raised by means of the background metric $g_{ij}$. 

By performing the Taylor expansion of $\sqrt{-\tilde g}f(\tilde R,\tilde K,\tilde\phi)$ around the background fields 
$\{g_{ij},\phi\}$ up to the second order in the perturbations, we get
\beq
I_\mathrm{E} [g,\phi]\sim-\frac{1}{16\pi G}\int\:d^4x\,\sqrt{-g}\: \aq f_0+{\cal L}_1+{\cal L}_2\, \cq \,,
\label{AAA3} \eeq
where, up to total derivatives, 
%
%
\begin{eqnarray}
{\cal L}_1 \Eqn{=} \frac14\,X\,h+f_{\phi}\phi\:, \\ 
{\cal L}_2 \Eqn{=} -\frac{1}{2}{f_R}h^i_k\nabla_i\nabla_jh^{jk}+\frac{1}{4}{f_R}h_{ij}\Delta h^{ij}
-\frac{1}{24}\,R{f_R}h_{ij}h^{ij}
+\frac{1}{2}{f_R}h\nabla_i\nabla_j h^{ij}
\nonumber\\&&
+{f_{R\phi}}{\ph}\nabla_i\nabla_j h^{ij}
+\frac{1}{2}{f_{RR}}\nabla_i\nabla_jh^{ij}\nabla_r\nabla_sh^{rs}
-{f_{RR}}\Delta h\nabla_i\nabla_j h^{ij}
-\frac{1}{4}\,R{f_{RR}}h\nabla_i\nabla_j h^{ij}
\nonumber\\&&
-\frac{1}{48}\,R{f_R}h^2
-\frac{1}{2}{f_K}{\ph}\Delta{\ph}
-\frac{1}{4}{f_R}h\Delta h
-{f_{R\phi}}h\Delta{\ph}
+\frac{1}{4}\,R{f_{RR}}h\Delta h
+\frac12\,{f_{\phi\phi}}{\ph}^2
\nonumber\\&&
-\frac{1}{4}\,R{f_{R\phi}}h{\ph}
+\frac{1}{32}\,R^2{f_{RR}}h^2
+\frac{X}{16}\,(h^2-2h_{ij}h^{ij})+\frac12\,{f_\phi}h{\ph}\:. 
\end{eqnarray}
Here, we have set $X=2f_0-Rf_R$. 
In this way, the on-shell Lagrangian density can directly be 
obtained in the limit $X\to0$ and $f_\phi\to0$. 

As is well known, it is convenient to carry out the standard expansion of the tensor field $h_{ij}$ in irreducible components~\cite{frad,perry,duff80}, 
namely, 
\beq
h_{ij}=\hat
h_{ij}+\nabla_i\xi_j+\nabla_j\xi_i+\nabla_i\nabla_j\sigma
+\frac14\,g_{ij}(h-\lap_0\sigma)\:, 
\label{tt}\eeq
where $\si$ is the scalar component, while $\xi_i$ and $\hat h_{ij}$ are 
the vector and tensor components with the properties
\beq
\nabla_i\xi^i=0\:,\hs\hs \nabla_i\hat h^i_j=0\:,\hs\hs \hat
h^i_i=0\:.
\label{AAA4} 
\eeq
In terms of the irreducible components of the $h_{ij}$ field, 
the one-loop contribution to the Lagrangian density, again disregarding 
the total derivatives, becomes 
\begin{eqnarray}
{\cal L}_2&=&
\frac1{32}\,\sigma\,\at
9{f_{RR}}\Delta\Delta\Delta\Delta
-3{f_R}{}\Delta\Delta\Delta
+6{f_{RR}}R\Delta\Delta\Delta 
\nn\cp\\&&\hs\hs\hs\ap 
-{f_R}R\Delta\Delta
+{f_{RR}}R^2\Delta\Delta
-3X\Delta\Delta
-RX\Delta\ct\,\sigma
\nonumber\\&&
{}+\frac1{32}\,h\,\at
9{f_{RR}}\Delta\Delta
-3{f_R}\Delta
+6{f_{RR}}R\Delta
-{f_R}^2R
+{f_{RR}}R^2
+{X}\ct\,h
\nonumber\\&&
{}+\frac1{16}\,h\,\at
-9{f_{RR}}\Delta\Delta\Delta
+9{f_R}\Delta\Delta
-6{f_R}\Delta\Delta
-6{f_{RR}}R\Delta\Delta
+{f_R}R\Delta
-{f_{RR}}R^2\Delta\ct\,\sigma
\nonumber\\&&
{}+\frac12\,\ph\,\at
-{f_K}\Delta+f_{\phi\phi}\ct\,\ph
+\frac14\,h\,\at
-3{f_{R\phi}}\Delta
+2f_\phi -{f_{R\phi}}R\ct\,\ph
\nn\\&&
{}+\frac14\,\sigma\,\at
+3{f_{R\phi}}\Delta\Delta
+{f_{R\phi}}R\Delta\ct\,\ph
\nonumber\\&&
{}+\frac1{16}\,\xi_i\,\at
4X\Delta
+4RX\ct\,\xi^i
+\frac1{24}\,\hat h_{ij}\,\at
6{f_R}\Delta
-{f_R}R
-3X\ct\,\hat h^{ij}\:.
\end{eqnarray}

Since the invariance under the diffeomorphisms renders the 
operator in the $(h,\si)$ sector not invertible, 
a gauge-fixing term and a corresponding ghost compensating term 
have to be added. 
We explore the class of gauge conditions, parameterized by the real parameter
$\rho$ as 
\beq 
\chi_k=\nabla_j h^j_k-\frac{1+\rho}4\,\nabla_k\,h\:. 
\nn\eeq 
This is the harmonic or the de Donder one 
corresponding to the choice $\rho=1$. 
As the gauge fixing, we choose the quite general term
\cite{buch} \beq {\cal
L}_{gf}=\frac12\,\chi^i\,G_{ij}\,\chi^j\,,\hs\hs
G_{ij}=\ga\,g_{ij}+\beta\,g_{ij}\lap\,, 
\label{AAA5} 
\eeq 
where the term proportional to $\ga$ 
on the right-hand side of the second equation 
is the one normally used in the Einstein gravity. 
The corresponding ghost Lagrangian reads~\cite{buch} 
\beq
{\cal L}_{gh}= B^i\,G_{ik}\frac{\de\,\chi^k}{\de\,\ep^j}C^j\,,
\label{AAA6} 
\eeq 
where $C_k$ and $B_k$ are the ghost and anti-ghost vector fields, 
respectively, while $\de\,\chi^k$ is the variation of 
the gauge condition due to an infinitesimal gauge transformation of 
the field. It reads 
\beq
\de\,h_{ij}=\nabla_i\ep_j+\nabla_j\ep_i\segue
\frac{\de\,\chi^i}{\de\,\ep^j}=g_{ij}\,\lap+R_{ij}+\frac{1-\rho}{2}\,\nabla_i\nabla_j\,.
\label{AAA7} 
\eeq 
Neglecting total derivatives, we get 
\beq 
{\cal L}_{gh}=B^i\,\at\ga\,H_{ij}+\beta\,\lap\,H_{ij}\ct\,C^j\,,
\label{AAA8} 
\eeq 
where we have set 
\beq
H_{ij}=g_{ij}\at\lap+\frac{R_0}{4}\ct+\frac{1-\rho}{2}\,\nabla_i\nabla_j\,.
\label{AAA9} 
\eeq
In irreducible components, we obtain
\begin{eqnarray}
{\cal L}_{gf} &=& \frac{\ga}2\aq\xi^k\,\at\lap_1+\frac{R_0}4\ct^2\,\xi_k
   +\frac{3\rho}{8}\,h\,\at\lap_0+\frac{R_0}3\ct\,\lap_0\,\si
\cp\nn\\&&\hs\ap
   {}-\frac{\rho^2}{16}\,h\,\lap_0\,h
-\frac{9}{16}\,\si\,\at\lap_0+\frac{R_0}3\ct^2\,\lap_0\,\si
\cq
\nn\\&&
{}+\frac{\beta}2\aq\xi^k\,\at\lap_1+\frac{R_0}4\ct^2\,\lap_1\xi_k
   +\frac{3\rho}{8}\,h\,\at\lap_0+\frac{R}4\ct\at\lap_0+\frac{R}3\ct\,
\lap_0 \si
\cp\nn\\&&\hs\ap
   {}-\frac{\rho^2}{16}\,h\,\at\lap_0+\frac{R_0}4\ct\,\lap_0 h
   -\frac{9}{16}\,\si\,\at\lap_0+\frac{R_0}4\ct\at\lap_0+
\frac{R_0}3\ct^2\,\lap_0 \si \cq\,, \label{AAA10} \eeq \beq {\cal
L}_{gh} &=& \ga\ag\hat B^i\at\lap_1+\frac{R_0}{4}\ct\hat C^j
+\frac{\rho-3}{2}\,b\,\at\lap_0-\frac{R_0}{\rho-3}\ct\,\lap_0 c\cg
\nn\\&&\hs {}+\beta\ag\hat
B^i\,\at\lap_1+\frac{R_0}{4}\ct\,\lap_1\,\hat C^j \cp
\nn \\
 &&\hs\hs\ap {}+\frac{\rho-3}{2}\,b\,\at\lap_0+\frac{R_0}{4}\ct
  \at\lap_0-\frac{R_0}{\rho-3}\ct\,\lap_0 c\cg\,,
\eeq
where the ghost irreducible components are defined by
\beq
C_k &\equiv& \hat C_k+\nabla_k c\,,\hs\hs \nabla_k\hat C^k=0\,,
\nn\\
B_k &\equiv& \hat B_k+\nabla_k b\,,\hs\hs \nabla_k\hat B^k=0\,,
\label{AAA11} 
\eeq
and for clarity, from now on we use the notation 
$\lap_0$, $\lap_1$, and $\lap_2$ for the Laplace-Beltrami operators
acting on scalars, traceless-transverse vector fields, and
traceless-transverse tensor ones, respectively. 

In order to compute the one-loop contribution to the effective action, 
we have to analyze the path integral for the bilinear part 
of the total Lagrangian 
\beq 
{\cal L}= {\cal L}_2+\,{\cal L}_{gf}+{\cal L}_{gh} \,,
\label{AAA12} 
\eeq
and take into account 
the Jacobian due to the change of variables with respect to 
the original ones. 
In this way, we get~\cite{buch,frad}
\beq
Z^{(1)}=e^{-\Ga^{(1)}}&=&\at\det G_{ij}\ct^{-1/2}\,\int\,D[h_{ij}]D[C_k]D[B^k]\:
\exp\,\at -\int\,d^4x\,\sqrt{g}\,{\cal L}\ct
\nn\\
&=&\at\det G_{ij}\ct^{-1/2}\,\det J_1^{-1}\,\det J_2^{1/2}\,
\nn\\&&\hspace{-1cm}
 \times \int\,D[h]D[\hat h_{ij}]D[\xi^j]D[\si] D[\hat C_k]D[\hat B^k]D[c]D[b]\:
    \exp\, \at-\int\,d^4x\,\sqrt{g}\,{\cal L}\ct\,. 
\eeq 
Here, $J_1$ and $J_2$ are the Jacobians due to the 
change of the variables in the ghost and tensor sectors, 
respectively~\cite{frad}, described by 
\beq 
J_1=\lap_0\,,\hs\hs
J_2=\at\lap_1+\frac{R_0}{4}\ct\at\lap_0+\frac{R_0}{3}\ct\,\lap_0\,,
\label{AAA13} \eeq and the determinant of the operator  $G_{ij}$,
acting on vectors, can be written as \beq \det G_{ij}=\mbox{const}\
\det\at\lap_1+\frac{\ga}{\beta}\ct\,\det\at\lap_0+\frac{R_0}4+\frac{\ga}{\beta}\ct\,,
\label{AAA14} 
\eeq
while it is trivial in the case $\beta=0$.

\section{One-loop effective action}


Now, a straightforward computation leads to the approximate effective action 
with the one-loop quantum correction. 
This is a quite complicated gauge-dependent quantity. 
For simplicity, and since we are mainly interested 
in the Landau gauge, we restrict our investigations to the class 
of gauges with an arbitrary parameter $\ga$ and fixed ones as 
$\rho=1$ and $\beta=0$. 
In this way, we have the formal equations
\beq
\Ga=I_\mathrm{E} [g,\phi]+\Ga^{(1)}\,,\hs\hs
   I_\mathrm{E} [g,\phi]=\frac{24\pi f_0}{GR^2}\,,
\eeq
\beq
\Ga^{(1)}&=&\frac12\,\ln\det\at b_4\lap_0^4+b_3\lap_0^3+b_2\lap_0^2+b_1\lap_0+b_0\ct
                                 -\ln\det\at-\lap_0-\frac{R}2\ct
\nn\\&&
                       {}+\frac12\,\ln\det\at-\lap_1-\frac{R}{4}-\frac{X}{2\ga}\ct
                                      -\ln\det\at-\lap_1-\frac{R}4\ct 
\nn\\&&
                     {}+\frac12\,\ln\det\at-\lap_2+\frac{R}{6}+\frac{X}{2f_R}\ct\,,
\eeq
where the coefficients $b_k$ are complicated expressions depending on the
function $f(R,K,\phi)$ and its derivatives. These are explicitly written 
in Appendix A, where the expressions for the simpler case of $f(R)$ are also 
presented. 
The determinant of a differential operator can be well defined by means of 
zeta-functions~\cite{report}, that on $S^N$ can be expressed in terms of 
the Hurwitz zeta functions as is explained in Appendix~\ref{A:det}.

The ``on-shell'' contribution is obtained 
in the limit $X\to0$ and $f_\phi\to0$, and 
as is well known, it is gauge independent. 
For the one-loop contribution, we find 
\beq
\Ga^{(1)}_\mathrm{on-shell}&=&\frac12\,\ln\det \aq \frac1{\mu^4}\, \at a_2\lap_0^2+a_1\lap_0+a_0\ct\cq
\nn\\&&
     {}-\frac12\,\ln\det \aq \frac1{\mu^2}\,\at-\lap_1-\frac{R}4\ct \cq
       +\frac12\,\ln\det\ \aq\frac1{\mu^2}\,\at-\lap_2+\frac{R}6\ct\cq\,.
\label{AAA16}
\eeq
Also the coefficients $a_k$ depend on the function $f$ and its derivatives and are written in Appendix A. 
Here, an arbitrary renormalization parameter $\mu^2$ has 
been introduced for dimensional reasons. 
Furthermore, we should mention another delicate point. 
The Euclidean gravitational action is not bounded from below 
due to the presence of $R$, 
because arbitrary negative contributions can be induced on $R$ by conformal 
rescaling of the metric. For this reason, 
we have also used the Hawking prescription to integrate over
imaginary scalar fields. Finally, the problem of presence of 
additional zero modes introduced by the decomposition in Eq.~(\ref{tt}) can 
be treated through the method proposed in Ref.~\cite{frad}. 

For physical applications, the most appropriate one is the Landau gauge~\cite{buch,frad,od}, that corresponds to the choice of gauge parameters 
$\rho=1$, $\beta=0$, and $\ga=\ii$. 
It is known that such a gauge condition in one-loop approximation makes the convenient effective action to be equal to the gauge-fixing independent effective action (for reviews, see~\cite{frad,buch}). 
In this case, we get 
\beq
\Ga^{(1)}_\mathrm{Landau} &=&
           \frac12\,\ln\det\aq\frac1{\mu^8}\,
           \at c_4\lap_0^4+c_3\lap_0^3+c_2\lap_0^2+c_1\lap_0+c_0\ct\cq
\nn\\&&
     {}-\frac12\,\ln\det \aq\frac1{\mu^2}\,\at-\lap_1-\frac{R}4\ct \cq
         -\ln\det \aq\frac1{\mu^2}\,\at-\lap_0-\frac{R}2\ct \cq
\nn\\&&
           {}+\frac12\,\ln\det\ \aq\frac1{\mu^2}\,\at-\lap_2-\frac{R}3+\frac{f_0}{f_R}\ct\cq\,.
\label{EA-G}
\eeq
The coefficients $c_k$ are also represented in Appendix~\ref{A:coeff}.

\section{Cosmological applications}

\subsection{Quantum equivalence between the Jordan and Einstein frames}

Modified gravity models are usually formulated in the so-called Jordan frame, 
where the gravitational Lagrangian density only depends on geometric 
invariants. However, for some models like $f(R)$ gravity, it is possible to 
formulate the theory in the so-called Einstein frame, where, by means of a suitable conformal transformation involving geometric quantities only, the original Lagrangian is replaced by the Einstein-Hilbert one plus the Lagrangian of a scalar field that explicitly takes account of an additional degree of freedom which presents in the original theory. 
At the classical level, the equivalence of the two formulations has been 
studied in many works (see, for instance, reviews~\cite{review}).
Here, we show that this is true also at the one-loop level 
for the on-shell effective actions (compare with corresponding one-loop equivalence of dilatonic gravity in different frames~\cite{no2000}). 
On the contrary, for the off-shell ones, the one-loop contributions 
are completely different. 

To this aim, we investigate $f(R)$ gravity in the Jordan and Einstein frames. 
The corresponding classical actions are represented as 
\beq
I_\mathrm{Jord}=\frac1{2\ka^2}\,\int\,d^4x\sqrt{-g}\,f(R)\,,\hs\hs\hs\mbox{for the Jordan frame,}
\eeq
\beq
I_\mathrm{Eins}=\frac1{2\ka^2}\,\int\,d^4x\sqrt{-\tilde g} \,\tilde f(\tilde R,K,\si)\,,\hs\hs\mbox{for the Einstein frame,}
\label{IEins}
\eeq
where
\beq
\tilde f(\tilde R,\tilde K,\si)= \tilde R-\frac32\,\tilde g^{ij}\,\partial_i\si\partial_j\si-V(\si)\,.
\eeq
In this section, all the quantities with the tilde are related to the metric $\tilde g_{ij}=e^\si g_{ij}$ in the Einstein frame. 
Also, note that $\si$ is not an arbitrary function, but it is related to $R$ 
as follows
\beq
e^{\si}=f'(R)\,,\hs\hs  R=\Phi(e^{\si})\,,\hs\hs \Phi\circ f'=1\,, 
\label{Jord-Eins}\eeq
where the prime denotes the derivative with respect to $R$. 
Moreover, the potential is implicitly defined by
\beq
V(\si) \equiv e^{-\si}\Phi(e^{\si})-e^{-2\si}f(\Phi(e^{\si}))\,.
\eeq
It is clear from Eq.~(\ref{AAA16}) that to verify the equivalence of the on-shell, one-loop effective actions corresponding to 
the classical actions above, it is sufficient to compare the corresponding scalar sectors. 

In the first case, the scalar contribution to the effective action can directly be read off from Eq.~(\ref{GafR}), that is, 
\beq
\Ga^\mathrm{Jord}_\mathrm{on-shell}&=&\frac12\,\ln\det\aq\frac1{\mu^2}\,\at-f_{RR}\at\lap_0+\frac R3\ct+\frac{f_R}3\ct\cq
\nn\\&&\hs\hs +\mbox{classical and higher spin contributions.}
\label{GaJ}
\eeq
In the second case, we have to examine Eq.~(\ref{AAA16}) and compute the coefficients $a_k$ by using the function $\tilde f(\tilde R,\tilde K,\si)$. We also 
take into consideration the fact that $\tilde\lap_0$ is related to 
$\lap_0$ via a conformal transformation. 
Consequently, we obtain 
\beq
\Ga^\mathrm{Eins}_\mathrm{on-shell}&=&\frac12\,\ln\det\aq\frac1{\tilde\mu^2}\,\at3\tilde\lap_0-V''(\si)\ct\cq
                \nn\\&&\hs\hs +\mbox{classical and higher spin contributions}
         \nn\\&=&\frac12\,\ln\det\aq\frac1{\tilde\mu^2}\,\at\frac{3\lap_0}{f_R}+\frac{R}{f_R}-\frac1{f_{RR}}\ct\cq
        \nn\\&&\hs\hs +\mbox{classical and higher spin contributions, }
\label{GaE}
\eeq
where the prime means the derivative with respect to $\sigma$, and 
the latter equality equivalent to Eq.~(\ref{GaJ}) 
with a trivial redefinition of $\tilde\mu$. 

As an example, we now explore $R^2$ gravity. In the Jordan frame, we have 
\beq
f(R)=R+\frac{R^2}{6M^2}\,,
\label{Starob-JF}
\eeq
with $M^2$ a mass parameter. 
For the Einstein gravity, there exists 
a new scalar degree of freedom, which is the so-called scalaron, 
as a consequence of the quadratic term in the classical action. 
It follows from Eq.~(\ref{BGF}) that there is the background solution of $R=0$,
and thus the related on-shell, one loop effective action reads
\beq
\Ga^\mathrm{Jord}_\mathrm{on-shell}&=&\frac12\,\ln\det \aq \frac1{\mu^2}\, \at-\lap_0+M^2\ct\cq
\nn\\&&\hs\hs +\mbox{classical and higher spin contributions.}
\label{sj}
\eeq
The scalaron itself manifests in the scalar functional determinant. 

In the Einstein frame, the new degree of freedom explicitly appears in the action as a scalar field with a suitable potential. In this particular case, 
we have 
\beq
\tilde f(\tilde R,\tilde K,\si)
    =\tilde R-\frac32\,\tilde g^{ij}\partial_i\si\partial_j\si-\frac32\,M^2\at1-e^{-\si}\ct^2
     =\tilde R-3\tilde K-\frac32\,M^2\at1-e^{-\si}\ct^2\,.
\eeq
It can directly be verified that 
the corresponding one-loop effective action is equivalent to the one in 
Eq.~(\ref{sj}), because $V''(0)=-3M^2$. 
It has to be noted that for such a simple model, 
the background solution corresponds to $\{R=0,\si=0\}$ 
as a consequence of Eq.~(\ref{BGF}). 

Through the replacement $\si\to \sqrt{2/3}\phi/M_\mathrm{P}$, we find 
the Lagrangian density in the standard form (see Appendix~\ref{A:staro}). 
In this case, 
the expression for the one-loop effective action is quite trivial, 
because the background geometry is flat ($R=0$). 
Therefore, we get the exact well-known result
\beq
\Ga^{(1)}_\mathrm{on-shell}=\frac{\cal V}{2}\,M^4\,\at\ln\frac{M^2}{\mu^2}-\frac32\ct\,,
\eeq
with ${\cal V}$ the (infinite) volume of the manifold.

\subsection{One-loop quantum-corrected $R^2$ gravity} 

In order to study the role of the one-loop quantum corrections to 
$R^2$ gravity, 
the off-shell one-loop effective action has to be used. 
The idea is to work in the Jordan frame and take the off-shell 
effective Lagrangian in the Landau gauge. 
Making use of Appendix~\ref{A:coeff}, 
for the model described by Eq.~(\ref{Starob-JF}) we get
\beq
\Ga^{(1)}_\mathrm{Landau} &=&\frac12\,{\cal V}\,M^2_\mathrm{P}\,\left(R+\frac{R^2}{6M^2}\right)
   +\frac12\,\ln\det\aq\frac1{\mu^2}\at-\frac12\,\lap_0-\frac{R}{2}\ct\cq
\nn\\&&\hs
     +\frac12\,\ln\det \aq\frac1{\mu^2}\,\at-\lap_1-\frac{R}4\ct \cq
         -\ln\det \aq\frac1{\mu^2}\,\at-\lap_0-\frac{R}2+M^2\ct \cq
\nn\\&&\hs\hs
           -\frac12\,\ln\det\ \aq\frac1{\mu^2}\,\at-\lap_2+\frac{R(R+12M^2)}{6(R+3M^2)}\ct\cq\,.
\label{EA-Gs}\eeq
Here, ${\cal V}=\frac{384\pi^2}{R^2}$ is the volume of $S^4$. 

For the classical description of inflation in the Jordan frame (see Appendix~\ref{A:staro}), the inflationary stage consists of two regimes. 
The first one is $M^2/R\ll1$, 
where the solution is a quasi de Sitter space-time, 
and the second one is $M^2/R\gg1$, in which 
the solution oscillates and inflation becomes over. 

In the first case (during inflation), by expanding Eq.~(\ref{EA-Gs}) for large $R$, we obtain
\beq
L(R)=\frac12\,M^2_\mathrm{P} \,\left[R+\frac{R^2}{6M^2}
      +\frac{R^2}{384\pi^2 M^2_\mathrm{P}}\,\left(C_1\,\ln\frac{R}{\mu^2}+C_2\right)\right]
      +O\left(\frac{M^2}{R}\right)\,,
\label{eq:5.13}
\eeq
where $C_1$ and $C_2$ are pure numbers, given by
\beq
C_1&=&\ze(0|-\hat\lap_1-3)-\ze(0|-\hat\lap_2+2)=O(1)\,,
\\
C_2&=&-\ze'(0|-\hat\lap_1-3)+\ze'(0|-\hat\lap_2+2)\sim300\,,
\eeq
with $\hat\lap$ 
the Laplace-Beltrami operator acting on the unitary hypersphere $S^4$
(see Appendix \ref{A:det}).

For the model described by the Lagrangian in Eq.~(\ref{eq:5.13}), 
by performing the conformal transformation as in Eq~(\ref{Jord-Eins}), 
we obtain the following action in the Einstein frame: 
\begin{equation}
I_\mathrm{Eins} = \frac1{\kappa^2}\,\int d^4 x \sqrt{-\hat{g}} \left(
\tilde{R} -\frac{3}{2} \tilde{g}^{ij} \partial_i 
\si\partial_j\si-V(\si)\right)\,, 
\label{eq:001}\end{equation}
where
\begin{eqnarray}
V(\sigma)=\at1-e^{-\sigma}\ct^2\,\,
    \frac{a+2b\ag 1+\log|e^{\sigma}-1|
    -\log\aq4|b\mu|W\left(\frac{|e^{\sigma}-1| e^{(a+b)/2b}}{4|b\mu|}\right)\cq\cg}
          {\left[4bW\left(\frac{|e^{\sigma}-1|]e^{(1+b)/2b}}{4|b\mu|}\right)\right]^2}\,,
\end{eqnarray}
with $W$ the Lambert function and 
\beq
a=\frac1{6M^2}+\frac{C_2}{384\pi^2M_P^2}\,,\quad\quad
b=\frac{C_1}{384\pi^2M_P^2}\,.
\eeq
As we said above, with the replacement $\si\to\sqrt{2/3}\phi/M_P$ we get the action in the standard form. 

The slow-roll parameters are pure numbers and so for 
their computation we can use units such that $1/2\kappa^2=M_P^2/2=1$. 
We acquire 
\beq
\ep=\at\frac1{V}\,\frac{dV}{d\phi}\ct^2=\frac13\,\at\frac{V'(\si)}{V(\si)}\ct^2\,,\quad\quad
\eta=\frac{2}{V}\,\frac{d^2V}{d\phi^2}=\frac{2}{3}\,\frac{V''(\sigma)}{V(\sigma)}\,.  
\eeq
Finally, the spectral index $n_{\mathrm{s}}$ and the tensor-to-scalar ratio 
are expressed as~\cite{Mukhanov:1981xt, L-L}
\begin{eqnarray}
n_{\mathrm{s}} = 1 -6\epsilon + 2\eta \,,\quad\quad r = 16 \epsilon \,. 
\label{eq:011}
\end{eqnarray}
In order to execute numerical calculations, now we choose $M\sim0.1\,M_P$, $\mu\sim M$, and $\phi_k \equiv\phi(\tilde t_k)\sim7.756 M_P$, where $\tilde t_k$ is the time when the density perturbation with 
a given scale $k$ first crosses the horizon $k/\left(\tilde a\tilde{H}\right)=1$. Here, $\tilde t$ is time in Einstein frame and 
$\tilde a$ the scale factor. 
With such values for the parameters, we find 
$n_s\sim 0.968$ and $r=0.0028$, which are practically the same as 
those obtained for the classical Starobinsky model. In fact, such values are essentially determined by the huge value of $\phi_k$ independently of the other parameters. 
Thus, what we could observe here is that the quantum gravity corrections might be small in terms of the values of $n_{\mathrm{s}}$ and $r$. 
It is clearly seen that 
the value of $n_{\mathrm{s}}$ could be compatible with the Planck result of 
$n_{\mathrm{s}} = 0.9603 \pm 0.0073\, (68\%\,\mathrm{CL})$~\cite{Ade:2013lta, Ade:2013uln}, 
whereas that of $r$ is out of the 1 $\sigma$ error range 
of the BICEP2 consequence 
$r = 0.20_{-0.05}^{+0.07}\, (68\%\,\mathrm{CL})$~\cite{Ade:2014xna}. 
Note, however, that we have calculated quantum corrections in the de Sitter 
background. Furthermore, owing to very complicated form of quantum 
corrections expressed in terms of zeta-functions, 
the expansion of the one-loop effective action has been made to get 
the expression in Eq.~(\ref{eq:5.13}). 
It could be considered that 
the calculation of even the one-loop quantum corrections in the 
most general background may bring new qualitative terms 
changing the above values of $n_{\mathrm{s}}$ and $r$. 

In the second case (at the end of inflation), taking the opposite limit
$M^2/R \gg 1$ in (\ref{EA-Gs}) and using equations in Appendix \ref{A:det}, we have the one-loop effective Lagragian with the Coleman-Weinberg quantum correction, namely, 
\beq
L(R)&=&\frac12\,M^2_P\,\left[R+\frac{R^2}{6M^2}
    -\frac{M^4}{32\pi^2 M^2_\mathrm{P}}\left(\ln\frac{M^2}{\mu^2}-\frac{3}{2}\right)
+O\at R^2\ln\frac{R}{M^2_\mathrm{P} \mu^2}\ct\right]
\nn\\&\sim&\frac12\,M^2_\mathrm{P} F(R)\,.
\label{q}
\eeq
We also note that here, there is two natural scales: the Planck scale 
$M^2_\mathrm{P}$ and the mass $M^2$, while 
 the effective Lagrangian explicitly depends on the scale parameter $\mu^2$.
Moreover, the effective cosmological constant 
\beq
\La(\mu)=\frac{M^4}{16\pi^2 M^2_\mathrm{P}}\,\left(\ln\frac{M^2}{\mu^2}-\frac{3}{2}\right)\,,
\label{cc}
\eeq
is positive or negative according to whether $\mu^2$ is smaller or larger 
than $M^2e^{-3/2}$. 
It vanishes for $\mu^2=M^2e^{-3/2}$, namely,  
the leading term of the quantum correction is absent at this scale.

The modified gravity $L(R)$ Lagrangian may be studied in the Einstein frame, where there are the Einstein gravity plus a scalar field $\phi$ (i.e., inflaton), 
that is induced by a conformal transformation related to the Ricci scalar
by 
\beq
R=3M^2 \left(e^{\sqrt{\frac23}\,\frac{\phi}{M_\mathrm{P}}}-1\right)\,.
\eeq 
We explore the quantum corrections only in the second regime, relevant for the end of inflation. 
In such a case, the inflaton potential reads 
\beq
V(\phi)=\frac12\,M^2_\mathrm{P}\left[\frac{3M^2}{2}\left(1-e^{-\sqrt{\frac23}\,\frac{\phi}{M_\mathrm{P}}}\right)^2
            +2\La(\mu)\,e^{-2\sqrt{\frac23}\,\frac{\phi}{M_\mathrm{P}}}\right]\,.
\label{eq:5.30}
\eeq
This potential has a minimum at $\phi=\phi_*$ defined by
\beq
e^{-\sqrt{\frac23}\,\frac{\phi_*}{M_\mathrm{P}}}
=1+\frac{4\La(\mu)}{3m^2}\,,
\eeq
namely, its value is slightly different from zero and given by
\beq
V(\phi_*)=\frac{3M^2_\mathrm{P} M^2\La(\mu)}{3m^2+4\La(\mu)}\,.
\eeq
Thus, for large $\phi$, the potential does not differ from that for the Starobinsky inflation, but near the minimum, there is a very small difference. 
This type of potential may be examined along the standard approach. 
The quantum corrections give contributions of the order $(M/M_\mathrm{P})^2$.

\subsection{Stability issues}

Equation~(\ref{AAA16}) is relatively simple also for quite general models, and 
 it may be used to investigate the stability of de Sitter background with respect to arbitrary perturbations. For this reason, we require 
the coefficients $a_0$, $a_1$, and $a_2$ to satisfy some constraints 
so that all the eigenvalues of the operators in Eq.~(\ref{AAA16}) can be non negative. 
The smallest eigenvalues of the Laplacian operators $-\lap_0$, $-\lap_1$, and $-\lap_2$ acting on scalar, vector, and tensor fields, respectively, are $0$, 
$R/4$, and $2R/3$. Thus, only the first term on the right-hand side of Eq.~(\ref{AAA16}) could be relevant to the stability problem. 
We here explore several particular cases. 
As a first example, we study the Einstein gravity with a cosmological constant.
In this case, by using $f=R-2\La$ and Eq.~(\ref{GafR}) 
in Appendix~\ref{A:coeff}, 
we obtain the well-known result~\cite{frad,perry,duff80}. 
For another approach to the same problem, see also Ref.~\cite{gianni}.

As a second example, we investigate a generalized model of 
\beq
f(R,K,\phi)=F(R)-\frac12\,g^{ij}\,\partial_i\phi\partial_j\phi=F(R)-K\,,
\eeq
where for convenience, we use units such that $M^2_P/2=1$. {}From 
Eqs.~(\ref{a0})--(\ref{a2}) in Appendix \ref{A:coeff}, we find 
\beq
f_K=-1\,,\hs a_0=0\,,\hs a_1=-F_R (RF_{RR}-F_R)\,,\hs a_2=-3F_RF_{RR}\,. 
\eeq 
In this case, what is related to the stability is the ratio $a_1/a_2$ that has to be non negative. 
The stability condition then becomes 
\beq
\frac{F_R}{F_{RR}}-R\geq0\,.
\label{m}
\eeq
Such a condition is exactly the same as the one obtained for the pure $F(R)$ case~\cite{cogno05,cogno06}. 

Another phenomenologically interesting example is 
the following scalar tensor model
\beq
f(R,K,\phi)=R-\frac12\,g^{ij}\,\partial_i\phi\partial_j\phi-\frac12\,m^2\phi^2+\xi R\phi^2\,.
\eeq
Here, 
\beq
f_R=1+\xi\phi^2\,,\quad f_K=-1\,,\quad f_\phi=(2\xi R-m^2)\phi\,,
\quad f_{\phi\phi}=2\xi R-m^2\,,\quad f_{R\phi}=2\xi\phi\,.
\eeq
On the background solution, $\phi$ and $R$ are constants, given by
\beq
R=\frac{m^2}{2\xi}\,, 
\quad 
\phi=\pm\frac{1}{\sqrt\xi}\,.
\eeq
As a consequence of the latter equation, we get $f_{\phi\phi}=0$. 
The coefficients $a_k$ read 
\beq
a_0=4m^2\,,\quad a_1=4(1+6\xi)\,,\quad a_2=0.
\eeq
If $\xi=-1/6$ (conformally invariant case), we have $a_1=0$. Hence, the bosonic sector disappears from the on-shell effective action, and eventually 
it plays no role. 
On the contrary, if $\xi\neq-1/6$, the stability of de Sitter solution is assured by condition 
\beq
-\frac{a_0}{a_1}=-\frac{m^2}{1+6\xi}\geq0\segue \xi<-\frac16\,.
\eeq
This means that the stable constant solution has the negative scalar curvature 
(i.e., the anti-de Sitter). 

In the general case, $a_2$ does not vanish, and the scalar contribution may be written in a factorized multiplicative form, neglecting the multiplicative anomaly, that can be ``absorbed'' in the $\mu^2$ parameter~\cite{eli1,eli2}. 
As a result, we acquire
\beq
\Ga_\mathrm{on-shell}&=&\frac{24\pi f_0}{GR^2}
           +\frac12\,\ln\det \aq \frac1{\mu^2}\, \at -\lap_0+X_1\ct\cq
            + \frac12\,\ln\det \aq \frac1{\mu^2}\, \at -\lap_0+X_2 \ct \cq
\nn\\&&\hs
     -\frac12\,\ln\det \aq \frac1{\mu^2}\,\at-\lap_1-\frac{R}4\ct \cq
       +\frac12\,\ln\det\ \aq\frac1{\mu^2}\,\at-\lap_2+\frac{R}6\ct\cq\,,
\label{AAA17}\eeq
where
\beq
X_{1,2}=\frac{1}{2}\at-\frac{a_1}{a_2} \pm \sqrt{\frac{a_1^2}{a_2^2}-\frac{4a_0}{a_2}}\ct\,.
\label{x}
\eeq
To have two positive roots, the following conditions have to be satisfied 
\beq
 \frac{a_1}{a_2}<0\,,\hs\hs
  \at\frac{a_1}{a_2}\ct^2\geq\frac{4a_0}{a_2}\geq0\,. 
\label{piu}
\eeq
Thus, it has been performed that using the one-loop effective action, 
we can analyze the stability of the maximally-symmetric background under consideration.

\section{Conclusions}

In the present paper, to solve issues on inflation, 
we have studied a generalized modified gravity model whose 
action is written by a generic function $f(R, K, \phi)$ of 
$R$, $K$ and $\phi$. We have explored the one-loop effective action in the de Sitter background both on-shell and off-shell in the Landau gauge. 
Also, we have investigated the stability of the de Sitter solutions and 
the one-loop quantum correction to inflation in $R^2$ gravity. 

Moreover, we have analyzed the spectral index $n_{\mathrm{s}}$ of scalar modes of the primordial density perturbations 
and those tensor-to-scalar ratio $r$, 
and make the comparison of the theoretical predictions with 
the observational data obtained by the Planck satellite as well as the BICEP2 experiments. 
Consequently, it has explicitly been shown that 
for sets of the wider ranges of the parameters, 
the value of $n_{\mathrm{s}}$ can explain the Planck analysis of 
$n_{\mathrm{s}} = 0.9603 \pm 0.0073\, (68\%\,\mathrm{CL})$, 
while the value of $r$ is not within the 1 $\sigma$ 
error range of the BICEP2 result 
$r = 0.20_{-0.05}^{+0.07}\, (68\%\,\mathrm{CL})$. 
For instance, 
$n_s\sim 0.968$ and $r=0.0028$ 
can be realized (when the values of model parameters are 
$M\sim0.1\,M_P$, $\mu\sim M$, and 
$\phi_k (=\phi(\tilde{t}_k)) \sim7.756 M_P$, 
as presented in Sec.~5.2). 
These resultant values mainly depend on the quite-large value of $\phi_k$, 
and therefore they are basically independent of the other model parameters. 
Since these results are similar to those in the Starobinsky inflation model, 
it is considered that quantum gravity might present only small corrections to 
the values of $n_{\mathrm{s}}$ and $r$. 

As a significant outlook, 
it should be emphasized that not only by the BICEP2 but also 
by other collaborations including B-Pol~\cite{B-Pol}, 
LiteBIRD~\cite{LiteBIRD}, POLARBEAR~\cite{POLARBEAR}, and QUIET~\cite{QUIET}, 
non-zero $r$ might be detected in the future. 
These data must present us some clues to understand high-energy physics 
describing the early universe.

\section*{Acknoweledgements}
The work has been supported in part by Russ. Min. of Education and Science, project TSPU-139 (S.D.O.) and the JSPS Grant-in-Aid 
for Young Scientists (B) \# 25800136 (K.B.).

\appendix
 
\section{Explicit calculations of the coefficients}
\label{A:coeff}

In this Appendix, we explicitly write down the coefficients which appear in the one-loop effective action for the general case $f(R,K,\phi)$. Moreover, 
we study the simpler case of $f(R)$, that cannot be obtained 
as a trivial limit of the general one. 

\ss{The general case: $f(R,K,\phi)$}

In such a case, we have
\begin{eqnarray}
a_0&=&f_R\,[Rf^2_{R\phi}+f_{\phi\phi}(f_R-Rf_{RR})]\,, \label{a0}\\
a_1&=& f_R\aq f_K(Rf_{RR}-f_R) +f^2_{R\phi}-3 f_{\phi\phi} f_{RR }\cq\,, \\
a_2&=& 3f_Kf_Rf_{RR}\,,
\label{a2}\end{eqnarray}
\begin{eqnarray}
b_0&=& \frac{4{f_{\phi}}^2 X}{\gamma}+4{f_{\phi}}^2 R
  -\frac{4{f_{\phi}}{f_{R\phi}}R X}{\gamma}-4{f_{\phi}}{f_{R\phi}}R^2
   +\frac{{f_{\phi\phi}}{f_{R}}R X}{\gamma}\nonumber\\&&
    {}+{f_{\phi\phi}}{f_{R}}R^2-\frac{{f_{\phi\phi}}{f_{RR}}R^2 X}{\gamma}
      -{f_{\phi\phi}}{f_{RR}}R^3-\frac{{f_{\phi\phi}}X^2}{\gamma}
       -{f_{\phi\phi}}R X\nonumber\\&&
      {}+\frac{{f_{R\phi}}^2 R^2 X}{\gamma}+{f_{R\phi}}^2 R^3\,, 
\nonumber \\
b_1&=& -\frac{{f_{K}}{f_{R}}R X}{\gamma}-{f_{K}}{f_{R}}R^2
        +\frac{{f_{K}}{f_{RR}}R^2 X}{\gamma}+{f_{K}}{f_{RR}}R^3\nonumber\\&&
          {}+\frac{{f_{K}}X^2}{\gamma}+{f_{K}}R X+\frac{4{f_{\phi}}^2{f_{R}}}{\gamma}
             -\frac{4{f_{\phi}}^2{f_{RR}}R}{\gamma}\nonumber\\&&
              {}+12{f_{\phi}}^2-\frac{12{f_{\phi}}{f_{R\phi}}X}{\gamma}
                 -20{f_{\phi}}{f_{R\phi}}R+\frac{2{f_{\phi\phi}}{f_{R}}X}{\gamma}
                 +4{f_{\phi\phi}}{f_{R}}R\nonumber\\&&
               {}-\frac{5{f_{\phi\phi}}{f_{RR}}R X}{\gamma}
               -7{f_{\phi\phi}}{f_{RR}}R^2-2{f_{\phi\phi}}X
          +\frac{5{f_{R\phi}}^2 R X}{\gamma}+7{f_{R\phi}}^2 R^2\,,
\end{eqnarray}
\begin{eqnarray}
b_2&=& -\frac{2{f_{K}}{f_{R}}X}{\gamma}-4{f_{K}}{f_{R}}R
                +\frac{5{f_{K}}{f_{RR}}R X}{\gamma}
                   +7{f_{K}}{f_{RR}}R^2\nonumber\\&&
                  {}+2{f_{K}}X-\frac{12{f_{\phi}}^2{f_{RR}}}{\gamma}-24{f_{\phi}}{f_{R\phi}}
                 +4{f_{\phi\phi}}{f_{R}}-\frac{6{f_{\phi\phi}}{f_{RR}}X}{\gamma}\nonumber\\&&
              {}-16{f_{\phi\phi}}{f_{RR}}R+\frac{6{f_{R\phi}}^2 X}{\gamma}+16{f_{R\phi}}^2 R\,,
\nonumber \\
b_3&=& -4{f_{K}}{f_{R}}+\frac{6{f_{K}}{f_{RR}}X}{\gamma}
+16{f_{K}}{f_{RR}}R-12{f_{\phi\phi}}{f_{RR}}+12{f_{R\phi}}^2\,,
\nonumber \\
b_4&=& 12{f_{K}}{f_{RR}}\,.
\end{eqnarray}
\beq
c_0&=& R\left[f_{\phi\phi}(2Rf_R-2f_0-R^2f_{RR})+(2f_\phi-Rf_{R\phi})^2\right]\,,\\
c_1&=& {f_0}\left(2{f_{K}}R-4{f_{\phi\phi}}\right) +12{f_\phi}^2-20{f_\phi}{f_{R\phi}}R\,,\nn\\&&
                       {}+R\left(-2{f_{K}}{f_{R}}R+{f_{K}}{f_{RR}}R^2+6{f_{\phi\phi}}{f_{R}}
                          -7{f_{\phi\phi}}{f_{RR}}R+7{f_{R\phi}}^2 R\right)\,, \\
c_2&=& 4{f_0}{f_{K}}-6{f_{K}}{f_{R}}R+7{f_{K}}{f_{RR}}R^2-24{f_\phi}{f_{R\phi}}
\nonumber \\  
&&           
{}+4{f_{\phi\phi}}\left({f_{R}}-4{f_{RR}}R\right)+16{f_{R\phi}}^2 R\,,\\
c_3&=& -4\left({f_{K}}\left({f_{R}}-4{f_{RR}}R\right)+3{f_{\phi\phi}}{f_{RR}}-3{f_{R\phi}}^2\right)\,,\\
c_4&=& 12{f_{K}}{f_{RR}}\,.
\eeq

\ss{The simpler case: $f(R)$}.

In this case, the one-loop effective action is quite simple as 
\beq
\Ga^{(1)}_\mathrm{on-shell}&=&\frac12\,\ln\det\aq\frac1{\mu^2}\,\at-f_{RR}\at\lap_0+\frac R3\ct+\frac{f_R}3\ct\cq
\nn\\&&
     {}-\frac12\,\ln\det \aq \frac1{\mu^2}\,\at-\lap_1-\frac{R}4\ct \cq
       +\frac12\,\ln\det\ \aq\frac1{\mu^2}\,\at-\lap_2+\frac{R}6\ct\cq\,.
\label{GafR}\eeq
\beq
\Ga^{(1)}_\mathrm{Landau} &=&
           \frac12\,\ln\det\aq\frac1{\mu^4}\,
           \at f_{RR}(6\lap_0^2+5R\lap_0-2f_R(\lap_0+R)+2f_0\ct\cq
\nn\\&&
     {}-\frac12\,\ln\det \aq\frac1{\mu^2}\,\at-\lap_1-\frac{R}4\ct \cq
         -\frac12\,\ln\det \aq\frac1{\mu^2}\,\at-\lap_0-\frac{R}2\ct \cq
\nn\\&&
           {}+\frac12\,\ln\det\ \aq\frac1{\mu^2}\,\at-\lap_2-\frac{R}3+\frac{f_0}{f_R}\ct\cq\,.
\label{GaLfR}\eeq

\section{$R^2$ gravity in the Jordan frame}
\label{A:staro}
The equation of motion for the classical $R^2$ gravity becomes 
\beq
\ddot H+\frac12\,M^2H+3H\,\dot H-\frac1{2H}\,\dot H^2=0\,.
\eeq
This equation is a highly non-linear second order differential equation, that has no a constant curvature solution $H=H_0$. 
However, we can find a solution of the form
\beq
H(t)=H(t_0)(1+v(t))\,,
\label{j}
\eeq
where $v(t)$ is a small quantity. 
At the first order in $v(t)$, we have the differential equation
\beq
\ddot v+3H(t_0)\dot v+\frac12\,M^2v=-\frac12\,M^2\,.
\label{j1}
\eeq
There may be two regimes. The first one is the early stage of the inflationary period $t_0=t_\mathrm{i}$, when
\beq
\frac{M^2}{H^2_\mathrm{i}}\ll1\,,\hs\hs H_\mathrm{i}=H(t_\mathrm{i})\,.
\eeq
If this condition is met, a solution that satisfies the initial 
conditions $v(t_\mathrm{i})=\dot v(t_\mathrm{i})=0$ is 
represented as 
\beq
H(t)=H_\mathrm{i}\left[1+\frac{p_2(1-e^{p_1(t-t_\mathrm{i})})-p_1(1-e^{p_2(t-t_\mathrm{i})})}{p_1-p_2} \right]\,,
\eeq
where
\beq
p_1=-\frac{3H_\mathrm{i}}{2}\left(1-\sqrt{1-\frac{2M^2}{H^2_\mathrm{i}}}\right)\,,\quad\quad
 p_2=-\frac{3H_\mathrm{i}}{2}\left(1+\sqrt{1-\frac{2M^2}{H^2_\mathrm{i}}}\right)\,.
\eeq
For this solution, we see that $R(t)\sim 12H^2_\mathrm{i}$.

The other interesting regime at the end of inflation ($t_0=t_\mathrm{e}$)
is obtained for
\beq
\frac{M^2}{H_\mathrm{e}^2}\gg1\,,\hs\hs H_\mathrm{e}=H(t_\mathrm{e})\,.
\eeq
The solution is an oscillating one, described by
\beq
H(t)=H_\mathrm{e}\left[1+e^{-3/2H_\mathrm{e} t}\sin\at\frac{M}{\sqrt{2}}\sqrt{1-\frac{18 H_\mathrm{e}^2}{M^2}}(t_\mathrm{e}-t)\ct\right]\,.
\eeq
Within this regime, $R$ is of the order of $12H_\mathrm{e}^2$ which is much smaller than $M^2$.

\section{Computation of the determinants}
\label{A:det}
In this Appendix, we recall some known facts about the evaluation of 
functional determinant of a differential 
elliptic non-negative operator $A$ defined on a compact $N$-dimensional
manifold without boundary (see, for example,~\cite{report}).
The starting point is the related zeta function
\beq
\zeta(s|A)=\sum_{n}\la_n^{-s}\,,\hs\hs\Re s>\frac{N}2\,,
\eeq
with $\la_n >0$ the non-vanishing eigenvalues of $A$.
The analytic continuation of the zeta-function is, under general conditions, 
regular at $s=0$. Thus, we may define
\beq
\ln\det\frac{A}{\mu^2} \equiv -\zeta'\at0\left|\frac{A}{\mu^2}\right.\ct\,,
\label{ZFdef}
\eeq
where the prime indicates derivative with respect to $s$.
Looking at Eq.~(\ref{EA-G}), we see that the one-loop 
effective action can be written in terms of the derivative of 
the zeta-functions corresponding to the Laplace-like operators 
acting on scalar, vector, and tensor fields on the 4-dimensional 
de Sitter space. In all of such cases, 
the eigenvalues of the Laplace operator are explicitly 
known and the zeta-functions can directly be computed by using
Eq.~(\ref{ZFdef}). 

The exact evaluation of the functional determinant is a difficult task 
even though one is dealing with constant curvature spaces. 
In our case, the functional determinant of the Laplace-like operators can be written as a finite sum of the Hurwitz zeta-functions, 
but for physical applications, approximate methods present acceptable results. 
For this reason, now we explore a method, 
that gives an analytic approximate expression for the determinant 
of an elliptic non-negative operator defined on a compact manifold 
$\cal M$ without boundary. 

In this work, we have to deal with the Laplace-like operators $L$ acting on scalar, vector, and tensor fields on the hyper-sphere $S^4$. They have the form
\beq
L=-\lap+E=\frac{R}{12}\,\hat L_u\,,\hs\hs \hat L_u=-\hat\lap+\frac{12}R\,E\,,
\eeq
where $E$ is a constant potential term, while $\hat\lap$ and $\hat L_u$ are 
operators acting on the unitary hyper-sphere $S^4$. 
It is also convenient to introduce the Laplace-like operator $\hat L$, given by
\beq
\hat L_u=\hat L+\al\,,\hs\hs\al=\frac{12}R\,E-\rho\,,
\label{LapE}
\eeq
with $\rho$ a pure number depending on the spin. 
The eigenvalues $\hat\la_n$ and their degeneration $d_n$ of $\hat L$ are 
well known. 
They can be represented in the form 
\beq
\la_n=(n+\nu)^n\,,\hs\hs d_n=c_1(n+\nu)+c_3(n+\nu)^3\,,
\eeq
so that the zeta-functions of $\hat L$ can trivially be expressed in term of the Hurwitz-zeta functions. 
In fact, we get
\beq
\ze(s|\hat L)=c_1\ze_\mathrm{H}(2s-1,\nu)+c_3\ze_\mathrm{H}(2s-3,\nu)\,,
\eeq
where $\nu,\rho,c_1,c_3$ depends on the spin according to the following table:
\beq
\hat L_0\segue &&\nu=\frac32\,,\hs\rho=\frac94\,\,\,\,,\hs c_1=-\frac1{12}\,\,\,\,, \hs c_3=\frac13\,,
\nn\\
\hat L_1\segue &&\nu=\frac52\,,\hs\rho=\frac{13}4\,,\hs c_1=-\frac9{4}\,\,\,\,\,\,\,, \hs c_3=1\,,
\nn\\
\hat L_2\segue &&\nu=\frac72\,,\hs\rho=\frac{17}4\,,\hs c_1=-\frac{125}{12}\,,\hs c_3=\frac53\,,
\eeq
As is well known, the zeta-function is related to the trace of the heat kernel
$K(t|A)=\Tr e^{-tA}$ via the Mellin transform. In particular, 
\beq
\ze(s|\hat L)=\frac1{\Ga(s)}\,\int_0^\ii\,dt\,t^{s-1}\,K(t|\hat L)\segue
K(t|\hat L)=\frac1{2\pi i}\,\int\,ds\,t^{-s}\Ga(s)\ze(s|\hat L)\,. 
\eeq
Integrating the last equation, we find the asymptotic expansion of $K(t|\hat L)$, that is, 
\beq
K(t|\hat L)\sim\sum_k \hat A_kt^{k-2}\,,\hs\hs \hat A_k=\Res\at\Ga(s)\ze(s|\hat L),s=2-k\ct\,.
\eeq
Clearly, the coefficients $\hat A_k$ depend on the spin, and 
they become 
\beq
&&
\hat A_0=c_3\,,\hs \hat A_1=c_1\,,\hs \hat A_2=c_1\ze_H(-1,\nu)+c_3\ze_H(-3,\nu)\,,
\nonumber \\
&&
\hat A_k=\frac{(-1)^k}{(k-2)!}\,\aq c_1\ze_H(3-2k,\nu)+c_3\ze_H(1-2k,\nu)\cq\,,\hs\hs \mathrm{for} \,\,\, k\geq3\,.
\eeq
Now, we have all the elements necessary to compute the zeta-function 
for the operator $\hat L_u$ in Eq.~(\ref{LapE}). 
Assuming $\al^2\gg1$, we find 
\beq
\ze(s|\hat L_u)&=&\frac1{\Ga(s)}\,\int_0^\ii\,dt\,t^{s-1}e^{-t\al}K(t|\hat L)
               \nn\\ &=&\frac1{\Ga(s)}\,\int_0^\ii\,dt\,t^{s-3}e^{-t\al}(\hat A_0+\hat A_1t+\hat A_2t^2)+G(s,\al)\,,
\eeq
where $G(s,\al)$ is an entire function of $s$, vanishing for $s\to0$ and for $\al\to\ii$. 
In this way, we get the asymptotic expression valid for large values of $\al$
\beq
\zeta(s|\hat L_u)=\frac{\hat A_0}{(s-1)(s-2)}\al^{2-s}+\frac{\hat A_1}{(s-1)}\al^{1-s}+\hat A_2\al^{-s}
                +\frac{1}{\Gamma(s)}\,O(\al^{-(1+s)})\,.
\label{voros}
\eeq
As the first consequence, we have the results
\beq
\zeta(0|\hat L_u)=\frac{\hat A_0}{2}\al^{2}+\hat A_1 \al+\hat A_2\,,
\label{0}
\eeq
\beq
\zeta'(0|\hat L_u)=-\frac{\hat A_0}{2}\,\al^2\,\at\ln\al^2-\frac32\ct
            +\hat A_1\al\,\at\ln\al -1\ct-\hat A_2\,\ln\al+O(\frac{1}{\al})\,, 
\eeq
and within this approximation valid for $\frac{E}{R}\gg1$, 
the regularized functional determinant for the original operator in Eq.~(\ref{LapE}) reads
\beq
\ln\det\frac{L}{\mu^2}&=& \nonumber
-\zeta'\at0\left|\frac{L}{\mu^2}\right.\ct=-\zeta'(0|\hat L_u)+\zeta(0|\hat L_u)\ln\frac{R}{12\mu^2}
\\
&=&\frac{A_0}{2}\,E^{2}\at \ln \frac{E}{\mu^2}-\frac32\ct
               -A_1E\,\at\ln \frac{E}{\mu^2}-1\ct + A_2\,\ln \frac{E}{\mu^2}+ O(\frac{1}{E})\,. 
\label{logDet}
\eeq
Here, the coefficients $A_k$ are related to $\hat A_k$, given by 
\beq
A_0=\frac{72\hat A_0}{R^2}\,,\hs A_1=\frac{12(\hat A_1+\ro\hat A_0)}{R}\,,\hs
A_2=\hat A_2+\ro\hat A_1+\frac12\,\ro^2\hat A_0\,.
\eeq
Using the above results, we can obtain the explicit expression of the one-loop effective action for the generalized gravity under consideration.


\end{document}